\documentclass[a4paper,11pt]{article}
\pdfoutput=1 
\usepackage{jcappub}
\usepackage{amsmath}
\usepackage{graphicx}
\usepackage{latexsym}
\usepackage{xspace}
\usepackage{color}
\usepackage{hyperref} 
\usepackage{bm}
\usepackage{relsize}
\usepackage{tabularx}
\usepackage{multirow}
\usepackage{amssymb}
\usepackage[table]{xcolor}
\usepackage{slashbox}
\usepackage{braket}
\usepackage{comment}
\usepackage[utf8]{inputenc}
\usepackage{slashed}
\usepackage{empheq}
\usepackage[T1]{fontenc} 

\allowdisplaybreaks

\makeatletter
\gdef\@fpheader{}
\g@addto@macro\bfseries{\boldmath}
\makeatother

\makeatletter
\renewcommand*\env@matrix[1][\arraystretch]{%
  \edef\arraystretch{#1}%
  \hskip -\arraycolsep
  \let\@ifnextchar\new@ifnextchar
  \array{*\c@MaxMatrixCols c}}
\makeatother

\newcommand{\ds}{\displaystyle}

\newcommand{\ie}{\textsl{i.e.~}}

\newcommand{\dd}{\mathrm{d}}

\newcommand{\sss}[1]{{\scriptscriptstyle{#1}}}

\newcommand{\uPl}{\mathrm{Pl}}

\newcommand{\usssPl}{\sss{\uPl}}

\newcommand{\Mp}{M_\usssPl}

\newcommand{\beq}{\begin{equation}}
\newcommand{\eeq}{\end{equation}}
\newcommand{\bea}{\begin{eqnarray}}
\newcommand{\eea}{\end{eqnarray}}

\newlength{\wsingfig}
\newlength{\wdblefig}
\newlength{\wquadfig}
\newlength{\wtriplefig}
\newcommand{\N}{\mathcal N}

\newcommand{\stsr}{_{ij}}                   
\newcommand{\ustsr}{^{ij}}                  
\newcommand{\Tsr}{_{IJ}}                    
\newcommand{\uTsr}{^{IJ}}                   

\subheader{}

\title{The separate universe approach with non minimal coupling to gravity}

\author[a]{Hugo Holland}
\emailAdd{hugo.holland@universite-paris-saclay.fr}

\author[a]{Julien Grain,}
\emailAdd{julien.grain@universite-paris-saclay.fr}

\affiliation[a]{Universit\'e Paris-Saclay, CNRS, Institut d'Astrophysique Spatiale, 91405, Orsay, France}

\date{today}

\begin{document}
\sloppy

\abstract{In this paper we extend a study of the validity conditions of the separate universe approach of cosmological perturbations to models of inflation with multiple fields that are non minimally coupled to gravity. The separate universe approach effectively describes the universe as a collection of homogeneous and isotropic patches, leading to an effective description of cosmological perturbations at large scales. This approximation is a necessary step in stochastic inflation and in the $\delta \N$ formalism. We study and compare the dynamics of cosmological perturbation theory at large scales to the dynamics derived in the separate universe picture for non linear sigma models where we have allowed a non minimal coupling between the fields and gravity. We found that the separate universe approach adequately reproduces the dynamics of the large scale perturbations, provided the wavelength of the modes considered is greater than a list of lower bounds derived from the background evolution of the model we consider. These lower bounds are then compared to those derived in the classically equivalent Einstein frame description of the Jordan frame model at hand. In particular these conditions are derived and compared for a Higgs-like toy model of inflation in the large field limit.}

\keywords{cosmological perturbation theory, inflation}

\maketitle

\flushbottom
\section{Introduction}
Inflation is the favoured model to describe the dynamics of the early universe \cite{Guth:1980zm, linde2005particlephysicsinflationarycosmology, Linde_2006}. This epoch of near exponential growth seeds the anisotropies in the cosmic microwave background and large scale structures of the universe. These anisotropies are predicted to be adiabatic, Gaussian distributed and with a near scale invariant power spectrum. Current observations agree with these predictions \cite{Planck:2018jri} and can be accurately modelled with inflation models driven by a single field. However current cosmological probes only measure a fraction of the $40$ to $60$ e-folds we expect inflation to have lasted, thus leaving room for a more complicated setup than the classical single field. On top of that, high energy constructions of inflationary models often require multiple scalar fields with derivative couplings, which in turn leads to a plethora of interesting phenomenology, such as primordial non gaussianities \cite{Bjorkmo:2019qno,Achucarro:2019lgo,achucarro2019orbitalinflationinflatingangular,Ach_carro_2020,Tzavara:2013wxa,SGrootNibbelink_2002,Garcia_Saenz_2018,Fumagalli_2019,Garcia_Saenz_2020,Pinol:2021aun} or the formation of Primordial Black Holes \cite{Fumagalli:2020adf,Geller:2022nkr,Qin:2023lgo}. 

Many models have been put forward in order to clearly define the field content of the universe at such an early time. One particularly interesting path is to only consider fields that are present in the standard model of particle physics. For example, it was proposed in 2007 to use the Higgs boson as the inflaton field \cite{BEZRUKOV2008703, Bezrukov_2011, Bezrukov_2014, Ezquiaga_2018, Fumagalli:2019ohr}. This idea remains a competitive model today since it effectively mimics Starobinsky inflation \cite{STAROBINSKY198099, STAROBINSKY1982175}, one of the models in strongest agreement with current observational data. Higgs driven inflation has been studied multiple times since this first attempt. In some multifield model of inflation, we can choose to minimally couple the fields to gravity, or add a coupling that depends explicitly on all fields. The former case is referred to as the Einstein frame, and the latter the Jordan frame \cite{Fujii:2003pa, Faraoni:2004pi}. Classically, a model described in the Jordan frame can always be reframed to the Einstein frame via a conformal transform \cite{Kaiser_2010}. This redefines the geometry of spacetime, as well as the geometry of the field space since the coupling metric between all fields is modified by this transformation. The correspondence between these two frames is particularly useful for Higgs inflation models since they are naturally described in the Jordan frame through the coupling between the Higgs and gravitational fields. However, they are much easier to study in the Einstein frame where large field and small field limits are well suited. This correspondence is also useful for models inspired by supersymmetry \cite{Geller:2022nkr,Qin:2023lgo}.

The seeds of large scale anisotropies are described in inflationary models as small deviations from a homogeneous and isotropic universe. These perturbative methods can be greatly simplified if we assume that the perturbed universe can be effectively described at large scales by a collection of homogeneous and isotropic patches. This so-called separate universe approach (SUA)\footnote{Or quasi isotropic picture, or separate universe approximation.} offers a simple way to compute cosmic inhomogeneities at large scales \cite{Wands_2000,Tanaka_2021,PhysRevD.68.103515,PhysRevD.68.123518,LIFSHITZ1992493,article,PhysRevD.49.2759,I_M_Khalatnikov_2002}, including non linear dynamics, by rephrasing the problem to a homogeneous and isotropic problem with different initial conditions in each patch. This method has been used in order to define several mathematical approaches to inflation, namely the $\delta \mathcal N$ formalism \cite{PhysRevD.42.3936,Sasaki_1996,Lyth_2005,PhysRevLett.95.121302,Sugiyama_2013,Tanaka:2024mzw,Artigas:2024xhc, Tada_2017,achucarro2019orbitalinflationinflatingangular}, stochastic inflation \cite{STAROBINSKY1982175,10.1007/3-540-16452-9_6,10.1143/PTP.80.1041, Pinol_2019,Pinol:2020cdp}, or their combination, the stochastic $\delta \mathcal N$ formalism \cite{Fujita:2013cna,Vennin:2015hra}. Since the SUA is an effective theory of large scales, we need to check that it indeed recovers the full dynamics and does not loose physical information. As such, the scale at which we can define the different patches is a key parameter and should not be chosen lightly. Such analysis have been conducted previously, both in single field models and multiple field models, as well as in the slow roll case and beyond \cite{Nambu_2005,doi:10.1142/10953,Pattison:2019hef,Artigas_2022,Cruces_2023}. Through previous works, the conditions with which the SUA can be used in the Einstein frame are clear \cite{Grain_2026}, however these conditions could be modified by the inverse conformal transform, and lead to a different set of conditions on which the SUA is allowed in the Jordan frame. We thus set to find this new list of conditions in order to compare them in both frames. 

Since Higgs inflation is naturally described in the Jordan frame, but effectively studied by going to the Einstein frame, it is particularly suited for our analysis. We start by a brief reminder of how non linear sigma models function in both the Jordan and the Einstein frame, as well as how we go from one to the other in Sec. \ref{sec:multifield}. We then execute perturbation theory in the Jordan frame, initially without using the SUA, then by making that assumption in order to compare both results and derive the set of conditions under which the SUA is valid. We also give the same comparison in the Einstein frame in Sec. \ref{sec:pert theory}. We then study a Higgs-like two field toy model and derive explicitly the set of conditions in both frames in order to grasp a better understanding of how the SUA is affected by the change of frames in Sec. \ref{sec:Higgs}. Finally, we conclude in Sec. \ref{sec:conclusion}.

\section{Multifield formalism}\label{sec:multifield}

\subsection{Jordan and Einstein frames}
In this paper we shall consider non linear sigma models as a proxy for multifield inflation with derivative and non-linear couplings. To this effect we consider theories with $n$ fields $\phi^I$ (with $I = 1,...,n$). We split our analysis in two classes which are classically equivalent. On the one hand we consider fields that are non minimally coupled to gravity, and on the other hand those that are minimally coupled to gravity. Both of these choices have the advantage of giving us a fully covariant theory in the field space.

Let us give a brief review of the classical equivalence between the Jordan and Einstein frames. A complete review of this can be found in \cite{Kaiser_2010}. We start from the action for $n$ fields non minimally coupled to gravity \cite{thieman_book}:
\begin{equation}
    \tilde S = \int \dd^4x \sqrt{-\tilde g} \left[f(\phi^I)  \tilde{R} - \frac 1 2 \tilde G_{IJ}\partial_\mu\phi^I\partial^\mu\phi^J + \tilde V(\phi^I)\right]. \label{ActionJordan}
\end{equation}
With $\tilde g_{\mu\nu}$ the metric, and $\tilde g$ its determinant. The non-linear coupling of our fields to gravity is  $f(\phi^I)$, $\tilde{R}$ the Ricci scalar, $\tilde V(\phi^I)$ the potential and $\tilde G_{IJ}$ the coupling metric of our fields, which is often set to $\delta_{IJ}$ in this so called Jordan frame. We can raise and lower indices in the field space with this metric and its inverse $\tilde G^{IJ}$.

In order to go from the Jordan frame to the Einstein frame, we perform a conformal transformation of the metric:
\begin{equation}
    \tilde{g}_{\mu\nu}  \rightarrow  g_{\mu\nu} =  \Omega^2(\phi^I)\tilde{g}_{\mu\nu}\, , \, \text{with } \Omega^2(\phi^I) = \frac{2 f(\phi^I)}{\Mp^2} \, ,\label{EinsteinAction}
\end{equation}
where $\Mp$ is the reduced Planck mass. This leads to a minimally coupled action, and we refer to this as the Einstein frame action
\begin{equation}
    S_E = \int \dd^4x \sqrt{-g} \left[\frac {\Mp^2} 2R - \frac 1 2 G_{IJ}\partial_\mu\phi^I\partial^\mu\phi^J + V(\phi^I)\right] \, ,
\end{equation}
where the potential has been rescaled to $ V(\phi^I) \equiv \frac{\tilde V(\phi^I}{\Omega^4(\phi^I)}$.The Ricci scalar is also redefined after the conformal transformation since it depends on the metric. The new phase space coupling metric is
\begin{equation}
    G_{IJ} \equiv \frac{\Mp^2}{2f} \tilde G_{IJ} + \frac{3\Mp^2}{2f^2}\frac{\partial f}{\partial \phi^I}\frac{\partial f}{\partial \phi^J}. \label{eq:G E & J}
\end{equation}
Interestingly, a conformal transformation of the space time metric has lead to a disformal transformation of the coupling metric. Similarly, the redefinition of the field momenta is a disformal transformation, and we have
\begin{equation}
    \pi_J \equiv \tilde\pi_J + \frac 3 {f \tilde v} \tilde G^{IK}\frac{\partial f}{\partial \phi^I}\frac{\partial f}{\partial \phi^J} \tilde\pi_K. \label{eq:pi E & J}
\end{equation}

The equations of motion derived in both frames are equivalent, provided that the reparametrisations are done accordingly. The observables one can predict with one frame or the other are also the same provided they are computed covariantly in the phase space. All classical quantities are unchanged by this reparametrisation, thus showing that the choice of frame is more a matter of convenience than a physical choice. For example computations are often easier in the Einstein frame, however the Jordan frame would be more appropriate for given models like Higgs inflation, where non minimal couplings emerge.

\section{Perturbation theory and the separate universe approach}\label{sec:pert theory}

In the context of stochastic inflation, both frames can be developed consistently, but their practical usefulness can depend on the implementation on the separate universe approach. This approximation consists in introducing small perturbations that are assumed to be homogeneous and isotropic as proxies for a complete perturbation theory approach at large scales. This is a way of quantising the fact that once a mode is stretched beyond the horizon, spatial gradients in its dynamics become negligible in front of time evolution. Thus one can describe each Hubble patch as an FLRW universe with different initial conditions for the background fields and energy densities. This approach is a necessary step in stochastic inflation or the stochastic $\delta \N$  formalism. Let us make the difference between the separate universe approach and a complete perturbation approach explicit. We follow a similar procedure to \cite{Grain_2026} where the computations were done in the Einstein frame, this time applied to the Jordan frame.

\subsection{Dynamics of perturbations}

We start by deriving the Hamiltonian associated to our action, before deriving the background equations of motion. On the one hand we linearise said Hamiltonian up to the second order, and on the other hand we linearise the background equations of motion in order to get the separate universe description. We will then be able to check when the separate universe approximation is valid, and we compare these validity conditions to those derived in \cite{Grain_2026} for the Einstein frame.

In order to go to a Hamiltonian formulation of general relativity, we need to foliate our four dimensional space time into three-dimensional space-like hypersurfaces with coordinate $\vec x$ and one time-like direction $\tau$. To do so, we introduce two Lagrange multipliers: $\tilde N(\tau,\vec x)$ the lapse function and $\tilde N_i(\tau,\vec x)$ the shift vector. We can now express the metric with these Lagrange multipliers in the ADM form
    \begin{equation}
        \dd s^2 = -\tilde N^2(\tau, \vec x) \dd\tau^2 + \tilde\gamma\stsr(\tau,\vec x)\left[\dd x^i + \tilde N^i(\tau,\vec x)\dd\tau\right]\left[\dd x^j + \tilde N^j(\tau,\vec x)\dd\tau\right]\,,
    \end{equation}
with $\tilde\gamma_{ij}$ the three dimensional metric on the hypersurfaces, and $\tilde\gamma^{ij}$ its inverse, which are used to lower and raise indices on these sheets. We now need to define the canonical momenta associated to each field and the three dimensional metric 
    \begin{equation}
        \tilde \pi_I\equiv \delta \tilde S/\delta\dot\phi^I \, ,\quad \tilde\pi^{ij}\equiv \delta \tilde S/\delta\dot{\tilde\gamma}_{ij} \, ,
    \end{equation}
where $\dot f\equiv\partial f/\partial \tau$ is the time derivative. We can now express the action in the Hamiltonian form
    \begin{equation}
    	\tilde S=\int \dd \tau\int\dd^3x\left(\pi_I\dot{\phi}^I+\tilde\pi^{ij}\dot{\tilde\gamma}_{ij}-\tilde N \tilde {\mathcal{C}}-\tilde N^i \tilde {\mathcal{D}}_i\right)\,, \label{eq:SJHam}
	\end{equation}
where $\tilde{\mathcal{C}}$ and $\tilde{\mathcal{C}}_i$ are the scalar and diffeomorphism constraints.
    \begin{align}
            \tilde{\mathcal C} =& \frac 1 {f \sqrt {\tilde\gamma}} [\tilde\pi\stsr\tilde\pi\ustsr - \frac 1 2 \tilde\pi^2] - f\sqrt{\tilde \gamma} \tilde R^{(3)} + \frac{1}{2 \sqrt{\tilde\gamma}} \tilde G\uTsr \tilde \pi_I\tilde\pi_J \\
            &+ \frac { \sqrt{\tilde \gamma }}{2}\tilde\gamma^{ij}\tilde G\Tsr  \partial_i\phi^I\partial_j\phi^J + \sqrt{\tilde \gamma}\tilde V \,, \label{scalar constraint}\\
            \tilde{\mathcal C}_i =& -2\partial_m(\tilde\gamma\stsr\tilde\pi^{jm}) + \tilde\pi^{mn}\partial_i\tilde\gamma_{mn} + \pi_I\partial_i\phi^I. \label{diff constraint}
    \end{align}
with $ \tilde\gamma$ the determinant of the spatial metric, $\tilde\pi\equiv \tilde \pi^{ij}\tilde\gamma_{ij}$ the trace of the gravitational momentum, and $ R^{(3)}$ the Ricci scalar of the hypersurfaces. From this expression of the action we get the hamiltonian $\tilde H\equiv \int \dd^3\vec x \left(\tilde N\tilde{\mathcal C} + \tilde N^i\tilde{\mathcal C}_i\right)$, from which the dynamics of the scalars and gravitational fields can be derived. As mentioned previously, the lapse and the shift are Lagrange multipliers, they do not have any associated momenta, thus do not take part in the dynamics. Varying the action with respect to them leads to
    \begin{equation}
        \tilde{\mathcal C} = 0 = \tilde{\mathcal C}_i.
    \end{equation}
The vanishing of the constraints holds throughout the evolution of our fields, ensuring that the solutions to the equations of motion stay on the surface of constraint in the field phase space. One can check that $\tilde{\mathcal C}$ and $\tilde{\mathcal C}_i$ are indeed first class constraints. From this we can define the background equations of motion and linearise our Hamiltonian up to the second order. As mentioned, we are following the same procedure than in previous works, we thus refer the reader to \cite{Grain_2026} for a more detailed derivation. 

Let us note that $(\phi^I,\pi_J)$ and $(\tilde\gamma^{ij},\tilde\pi_{mn})$ form canonical pairs, with the following Poisson brackets: $\{\phi^I(\tau,\vec x);\pi_J(\tau,\vec y)\} = \delta^I_J \delta^3(\vec x - \vec y) $ and $\{\tilde\gamma\stsr(\tau,\vec x),\tilde\pi^{mn}(\tau,\vec y )\} = \frac 1 2 (\delta^m_i\delta^n_j + \delta^n_i\delta^m_j)\delta(\vec x - \vec y)$. Now that we have defined the Hamiltonian, we can also write the time derivative of any function $h$ as
\begin{equation}
    \dot h(\phi^I,\tilde\pi_I;\tilde\gamma\stsr,\tilde\pi\ustsr) = \left\{ h(\phi^I,\tilde\pi_I;\tilde\gamma\stsr,\tilde\pi\ustsr),\tilde H[N, N^i]\right\}.
\end{equation}
Applying this to our fields we get the following equations of motion
    \begin{align}
        \dot \phi^I =& \frac {\tilde N}{\sqrt{\tilde\gamma}}\tilde G\uTsr\tilde \pi_J + \tilde N^i\partial_i\phi^I \, ,\label{eq:dotphigen} \\
        \tilde D_\tau  \tilde\pi_I =& - \sqrt{\tilde \gamma} \tilde N \frac{\partial \tilde V}{\partial\phi^I} - \frac{\partial f}{\partial \phi^I}\frac {\tilde N} {f^2 \sqrt {\tilde\gamma}} [\tilde\pi\stsr\tilde\pi\ustsr - \frac 1 2 \tilde\pi^2] + \tilde D_i (\sqrt{\tilde\gamma }\tilde N \tilde G\Tsr \tilde\gamma\ustsr \partial_j \phi^I) +  D_i(\tilde N^i\pi_I) \, ,
    \end{align}
     where $D_\mu$ is the covariant derivatives in the field space. Applied to vectors $U^I$ and covectors $W_I$, it reads
    \begin{equation}
        \tilde D_\mu U^I = \partial_\mu U^I + \tilde \Gamma^I_{LK}\partial_\mu\phi^LU^K \quad \mathrm{and} \quad \tilde D_\mu W_I = \partial_\mu - \tilde \Gamma^K_{IL}\partial_\mu\phi^LW_K\,,
    \end{equation}
    with $\tilde \Gamma^K\Tsr$ the Christoffel coefficients associated to the coupling metric $\tilde G\Tsr$. From now on, partial derivatives in the field space will be denoted by a comma \ie $U^J_{\,\,,I}$, and covariant derivatives  by a semicolon \ie $U^J_{\,\,;I}$. Similar expressions for covariant derivatives in the Einstein frame will be used in Sec. \ref{sssec:Einstein frame conditions}. However it is clear that the covariant derivative of a tensor defined in the Jordan frame is defined with this expression, whereas the covariant derivative of an Einstein frame tensor uses the Christoffel symbols of the Einstein frame coupling metric $G\Tsr$. For the gravitational degrees of freedom, we get the following time evolutions
    \bea
    	\dot{\tilde\gamma}_{ij}&=&\frac{\tilde N}{f\sqrt{\tilde \gamma}}\left[2\tilde \pi_{ij}-\tilde \gamma_{ij}\tilde \pi\right]+2\tilde \gamma_{mj}\partial_i\tilde N^m+\tilde N^m\partial_m\tilde \gamma_{ij}, \\
        \dot{ \tilde \pi}^{ij}&=&\frac{ \tilde N \tilde\gamma^{ij}}{f\sqrt{ \tilde\gamma}}\left( \tilde\pi_{mn} \tilde\pi^{mn}-\frac{1}{2} \tilde\pi^2\right)+\frac{\tilde N}{f\sqrt{ \tilde\gamma}}\left( \tilde\pi \tilde\pi^{ij}-2 \tilde\gamma_{mn} \tilde\pi^{im} \tilde\pi^{jn}\right)+\tilde N f\sqrt{ \tilde\gamma}\frac{\delta \tilde R^{(3)}}{ \delta \tilde\gamma_{ij}} \nonumber \\
	    &&+\frac{\tilde N}{2 \sqrt {\tilde\gamma}} \tilde\gamma^{ij} \tilde G\uTsr \pi_I\pi_J +\frac{\tilde N}{2}\tilde \gamma^{im} \tilde\gamma^{jn}\tilde G\Tsr  \partial_m\phi^I\partial_n\phi^J-2\tilde\pi^{jm}\partial_m \tilde N^i-\partial_m\left(\tilde N^m \tilde \pi^{ij}\right).
    \eea
Let us note that these equations have been simplified using $\tilde{\mathcal C} = 0$, but remain quite difficult to read.

We wish to apply this formalism to inflationary cosmology. To this effect all fields and the Lagrange multipliers will be split into a background quantity and a perturbation. Perturbations will be treated as small deviations from the background quantities.

\subsubsection*{Homogeneous and isotropic background}
To study the time evolution of the background quantities, we suppose we are working in a homogeneous and isotropic cosmology. Let us introduce two useful quantities, $\tilde v\equiv \tilde a^3$ and its conjugate momenta $\tilde \theta(\tau)$, where $\tilde a(\tau)$ is the usual scale factor. Note that these quantities are dependent on time only. This allows us to rewrite our line element as:
\beq
	\dd s^2=-\tilde N^2(\tau)\dd\tau^2+\tilde v^{2/3}(\tau)\delta_{ij}\dd x^i\dd x^j,
\eeq 
where we have also introduced $\delta_{ij}$ the time independent flat three dimensional metric. Here the lapse is also only dependent on time, and the shift disappears from the metric as it is an inherently anisotropic degree of freedom. We can now write the background constraints and the background equations of motion
\bea
	\dot\phi^I&=&\frac{\tilde N}{\tilde v}\tilde G^{IJ}\tilde \pi_J, \label{eq:dotphi} \\
	\tilde D_\tau \tilde\pi_I&=&-\tilde N\tilde v\tilde V_{;I} - \tilde N \frac{f_{;I}}{f^2} \frac{3\tilde v\tilde\theta^2}{8}, \label{eq:dotpi}
\eea
and 
\bea
	\dot{v}&=&-\frac{3\tilde N}{2\Mp^2}\tilde v\tilde\theta, \label{eq:dotgamma} \\
	\dot{\theta}&=&\tilde N\frac{\tilde G^{IJ}\tilde\pi_I\tilde\pi_J}{\tilde v^2}. \label{eq:dotpig}
\eea
Since we only have a scalar constraint here, these equations of motion have to be solved under the condition $\tilde{\mathcal C^{(0)}} = 0$. The constraint 
\begin{equation}
    \tilde{\mathcal C}^{(0)} \equiv \frac{G^{IJ}\tilde\pi_I\tilde\pi_J}{2v^2} -  \frac{3\theta^2}{8 f} + V = 0 \label{eq:C0}
\end{equation}
has in fact already been used here to simplify the equation of motion of $\theta$. We note that for a constant coupling, the second term in the right hand side of Eq. (\ref{eq:dotpi}) drops, and we recover results as derived in the Einstein frame.

\subsubsection*{Cosmological perturbation theory}
Now that we have defined our background equations we can define the small fluctuations around the solutions as follows
\bea
	\delta\phi^I(\tau,\vec{x})&=&\phi^I(\tau,\vec{x})-\phi^I(\tau)\,, \label{eq:deltaphi} \\
	\tilde{\delta\pi}_J(\tau,\vec{x})&=&\tilde\pi_J(\tau,\vec{x})-\tilde\pi_J(\tau)\,, \label{eq:deltapi} \\
	\tilde{\delta\gamma}_{ij}(\tau,\vec{x})&=&\tilde\gamma_{ij}(\tau,\vec{x})-\tilde v^{1/3}(\tau)\,\delta_{ij}\,, \\
	\tilde{\delta\pi}^{ij}(\tau,\vec{x})&=&\tilde\pi^{ij}(\tau,\vec{x})-\frac{1}{2}\tilde v^{1/3}(\tau)\theta(\tau)\,\delta^{ij}\,, \\
    \tilde{\delta N}(\tau,\vec{x})&=& \tilde N(\tau,\vec{x}) - \tilde N(\tau)\,, \\
    \tilde{\delta N}_i(\tau,\vec{x})&=& \tilde N_i(\tau,\vec{x}) - \tilde N_i(\tau).
\eea
Cosmological perturbations are often treated in the Fourier space since our background solutions are homogeneous and isotropic, thus enforcing that different Fourier modes decouple. Let us make this change. For any given tensor field $T_{i\cdots j}(\tau,\vec x)$, we define its Fourier transform as
\begin{equation}
    	T_{i\cdots j}(\tau,\vec{k})=\ds\int\frac{\dd^3\vec{x}}{(2\pi)^{3/2}}T_{i\cdots j}(\tau,\vec{x})e^{-i\vec{k}\cdot\vec{x}}\,,
\end{equation}
where $\vec k$ is the comoving wavevector. its indices are raised and lowered with the flat metric $\delta_{ij}$. Let us note that since our fields are real valued, their Fourier transforms need to verify $T_{i\cdots j}(\tau,\vec{k})^* = T_{i\cdots j}(\tau,-\vec{k})$. Cosmological perturbations are usually scalar, vectorial and tensorial, however we focus in this work solely on scalars since scalar, vector and tensorial modes decouple at linear order. Let us point out that there have been attempts at defining the SUA for non scalar degrees of freedom as well \cite{Tanaka_2021,Tanaka:2023gul,Tanaka:2024mzw}, but these go beyond the scope of our work. We can thus simplify the tensorial quantities as follows in Fourier space
\beq
	\tilde{\delta N}^i(\tau,\vec{k})=i(k^i/k)\tilde{\delta N_1}(\tau,\vec{k}),
\eeq
since the shift vector can be expanded as a curl-free part and a divergence-free vector which we set to zero. Similarly for the perturbations of the spatial metric and its momentum, we write
\bea
	\tilde{\delta\gamma}_{ij}(\tau,\vec{k})&=&\tilde{\delta\gamma}_1(\tau,\vec{k})M^1_{ij}(\vec{k})+\tilde{\delta\gamma}_2(\tau,\vec{k})M^2_{ij}(\vec{k}), \\
	\tilde{\delta\pi}_{ij}(\tau,\vec{k})&=&\tilde{\delta\pi}_1(\tau,\vec{k})M_1^{ij}(\vec{k})+\tilde{\delta\pi}_2(\tau,\vec{k})M_2^{ij}(\vec{k}).
\eea
where
\bea
	M^1_{ij}(\vec{k})=\frac{\delta_{ij}}{\sqrt{3}}&\,\,\,\mathrm{and}\,\,\,& M^2_{ij}(\vec{k})=\sqrt{\frac{3}{2}}\left(\frac{k_ik_j}{k^2}-\frac{\delta_{ij}}{3}\right).
\eea
We can now fully linearise our Hamiltonian up to the second order. The linear action will drop on the background solutions, and we are left in Fourier space with $\tilde S = \tilde S^{(0)}+\tilde S^{(2)}$, where
\bea
\label{second order action}	\tilde S^{(2)}&=&\ds\int\dd \tau\int_{\mathbb{R}^{3+}}\dd^3k\left[\left(\tilde{\delta\pi}^\star_I\dot{\delta\phi^I}+\mathrm{c.c.}\right)+\sum_{A=1}^2\left(\tilde{\delta\pi}^\star_A\dot{\tilde{\delta\gamma}}_A+\mathrm{c.c.}\right)\right. \\
	&&\left.-\left(\tilde{\delta N}^\star\,\tilde{\mathcal{C}}^{(1)}+\mathrm{c.c.}\right)+\frac{k^i}{k}\left(i\tilde{\delta N}^\star_1\,\tilde{\mathcal{D}}^{(1)}_i+\mathrm{c.c.}\right)-2\tilde N\tilde{\mathcal{C}}^{(2)}\right], \nonumber
\eea
where $\mathrm{c.c.}$ indicates a complex conjugate, which comes from the integral on $\mathbb{R}^{3+}$. Let us note that one needs to be careful to avoid double counting degrees of freedom when deriving this expression. However stopping here would not be satisfactory since the linearised Hamiltonian is not explicitly covariant in the field space despite starting from a fully covariant theory.

As first presented in \cite{Gong_2011} the perturbations of the fields and their momenta are not covariant under field redefinitions. This explains why a direct linearisation will lead to a non manifestly covariant hamiltonian. To this effect we introduce covariant variables that are intimately linked to the original perturbations. Up to the second order we have
\bea
	\delta\phi^I &=& {\tilde Q}^I -\frac 1 2  \tilde\Gamma^I_{LK} {\tilde Q}^L {\tilde Q}^K+\mathcal{O}(\lambda^3) \label{mapping phi main}, \\
            \tilde{\delta\pi}_I &=& \tilde P_I +  \tilde\Gamma^K\Tsr \tilde \pi_K  {\tilde Q}^J + \tilde \Gamma^K\Tsr \tilde P_K {\tilde Q}^J \label{mapping pi main} \\ \nonumber
            &&+ \frac 1 2 \left(\tilde \Gamma^S_{IJ,K} - \tilde \Gamma^S_{IR}\tilde \Gamma^R_{JK} + \tilde\Gamma^R\Tsr\tilde\Gamma^S_{RK}-\frac{1}{3}{ \tilde R_{IJK}}^S\right)\tilde\pi_S {\tilde Q}^J {\tilde Q}^K+\mathcal{O}(\lambda^3).
\eea
We can now proceed to a canonical transformation of the linearised hamiltonian to express it as a function of the new variables and we get the following manifestly covariant expression:
\bea
\label{second order action covariant}	S^{(2)}_J&=&\ds\int\dd \tau\int_{\mathbb{R}^{3+}}\dd^3k\left[\left(\tilde P^\star_I\tilde D_\tau {Q}^{I}+\mathrm{c.c.}\right)+\sum_{A=1}^2\left(\tilde{\delta\pi}^\star_A\dot{\tilde{\delta\gamma}}_A+\mathrm{c.c.}\right)\right. \\
	&&\left.-\left(\tilde{\delta N}^\star\,\tilde{\mathcal{C}}^{(1)}+\mathrm{c.c.}\right)+\frac{k^i}{k}\left(i\tilde{\delta N}^\star_1\,\tilde{\mathcal{D}}^{(1)}_i+\mathrm{c.c.}\right)-2\tilde N\tilde{\mathcal{K}}^{(2)}_{\mathrm{cov}}\right], \nonumber
\eea
with at first order, and defining $\tilde{\mathcal D}_i^{(1)} = k_i\tilde{\mathcal D}^{(1)}$
\begin{align}
	\tilde{\mathcal C}^{(1)}=&\ds \left(\tilde v \tilde V_{;I} + \frac{3f_{;I} \tilde v \tilde \theta^2}{8f^2}\right){\tilde Q}^I+\frac{1}{\tilde v}\tilde G^{KI}\tilde \pi_K \tilde P_I-\frac{\tilde v^{1/3}}{\sqrt{3}}\left[\frac{1}{2}\left(\tilde\rho+3\tilde p\right)+2 f\frac{k^2}{\tilde v^{2/3}}\right]\tilde{\delta\gamma}_1+\sqrt{\frac2 3 }f\frac{k^2}{\tilde v^{1/3}}\tilde{\delta\gamma}_2 \\
	\nonumber &\ds-\frac{\sqrt{3}}{2 f}\tilde v^{2/3}\tilde \theta\,\tilde{\delta\pi}_1  \,,\\
	\tilde{\mathcal D^{(1)}}=&\,\ds\tilde\pi_I{\tilde Q}^I+\frac{1}{\sqrt{3}}\tilde v^{1/3}\tilde\theta\left(\frac{1}{2}\tilde{\delta\gamma}_1-\sqrt{2}\tilde{\delta\gamma}_2\right)-\frac{2}{\sqrt{3}}\tilde v^{2/3}\left(\tilde{\delta\pi}_1+\sqrt{2}\tilde{\delta\pi}_2\right).\label{eq:constcov}
\end{align}
And the second order reads
\bea
    \tilde{\mathcal{K}}^{(2)}_{\mathrm{cov}} &=& \frac{\tilde v}{4}\left(\frac{k^2}{\tilde v^{2/3}}\delta_{IJ} + \tilde V_{;IJ} -\frac{1}{\tilde v^2}\tilde R_I{}^{KL}{}_J\bar\pi_K\bar\pi_L -\frac{3 \tilde v\tilde\theta^2}{8f}\left(\frac{f_{;I}f_{;J}}{f^2} - \frac{f_{;IJ}}{2 f}\right)\right)\left({\tilde Q}^I{\tilde Q}^{J\,\star}+ \mathrm{c.c.}\right) \\
    &&+\frac{1}{4\tilde v}\tilde G\uTsr\left(\tilde P_I \tilde P_J^\star+ \mathrm{c.c.}\right)  + \frac{\tilde v^{1/3}}{2f}\left(-|\tilde{\delta\pi}_1|^2 + 2|\tilde{\delta\pi}_2|^2\right)+\frac{1}{12\tilde v^{1/3}}\left(5\tilde \rho+3 \tilde p-\frac{2fk^2}{\tilde v^{2/3}}\right)|\tilde{\delta\gamma}_1|^2 \nonumber \\
	&&+ \frac{1}{12\tilde v^{1/3}}\left(5\tilde \rho+3\tilde p-\frac{2fk^2}{2\tilde v^{2/3}}\right)|\tilde{\delta\gamma}_2|^2+ \frac{\sqrt{2}\,fk^2}{12\tilde v}\left(\tilde{\delta\gamma}_1\tilde{\delta\gamma}_2^\star+\mathrm{c.c.}\right) \nonumber \\ 
	&&-\frac{\tilde\theta}{8f}\left(\tilde{\delta\pi}_1\tilde{\delta\gamma}_1^\star + \mathrm{c.c.}\right) + \frac{\tilde\theta}{4f}\left(\tilde{\delta\pi}_2\tilde{\delta\gamma}_2^\star + \mathrm{c.c.}\right)  + \frac{ f_{;I} }{4 \sqrt 6 f} \left( {\tilde Q}^I\tilde{\delta\gamma_2} ^\star+ \mathrm{c.c.} \right) \nonumber \\
    && +\left[ \frac{\sqrt{3}\tilde v^{1/3} \tilde V_{;I}}{4} + \frac{\sqrt 3 f_{;I} \tilde v^{1/3}\tilde \theta^2}{32 f} - \frac{ f_{;I} }{4 \sqrt 3 f} \frac{k^2}{\tilde v^{1/3}}\right]\left( {\tilde Q}^I\tilde{\delta\gamma_1} ^\star+ \mathrm{c.c.} \right)\nonumber \\
    && + \frac{\sqrt 3 f_{;I} \tilde v^{2/3}}{4 f}\left( {\tilde Q}^I\tilde{\delta\pi}_1 ^\star+ \mathrm{c.c.} \right) -\frac{\sqrt{3} \tilde G\uTsr\tilde \pi_J}{4 \tilde v^{5/3}}\left(P_I\tilde{\delta\gamma}_1 ^\star+ \mathrm{c.c.}\right) .\nonumber
\eea
The rest of the expressions in the Hamiltonian were either scalar or vector quantities in the field space, thus automatically covariant, and remain unchanged by the transformation. Let us also note that $\tilde H_{\text{cov}}[\tilde N,\tilde N^i]$ is not a hamiltonian in the usual sense but a \textit{covariant} hamiltonian since we have written the action with a covariant derivative of ${\tilde Q}^I$. The hamiltonian equation we get from this expression for a field space tensor $U_{I\cdots J}$ is thus
\begin{equation}
    \tilde D_\tau U_{I\cdots J} = \{U_{I\cdots J}, \tilde H_{\text{cov}}\}.
\end{equation}
Applying this to our covariant perturbations gives
\bea
	&&\left\{\begin{array}{rcl}
	\tilde D_\tau \tilde{\delta\gamma}_1&=&\ds\frac{-2}{\sqrt{3}}\tilde v^{2/3}k\delta \tilde N_1-\frac{\sqrt{3}}{2f}\tilde v^{2/3}\tilde\theta\delta \tilde N-\frac{\tilde N}{2f}\left(2\tilde v^{1/3}\tilde{\delta\pi}_1+\frac{\tilde\theta}{2}\tilde{\delta\gamma}_1\right) + \frac{\sqrt 3 \tilde f_{;I} \tilde v^{2/3}}{2 f} {\tilde Q}^I\\
	\tilde D_\tau \tilde{\delta\pi}_1&=&\ds\frac{-\tilde v^{1/3}\tilde\theta}{2\sqrt{3}}k\delta \tilde N_1+\frac{\tilde v^{1/3}}{\sqrt{3}}\left[\frac{1}{2}\left(\tilde\rho+3\tilde p\right)+2f\frac{k^2}{\tilde v^{2/3}}\right]\delta \tilde N \\
	&&\ds-\tilde N\left[\frac{1}{6\tilde v^{1/3}}\left(5\tilde\rho+3\tilde p-2f\frac{k^2}{\tilde v^{2/3}}\right)\tilde{\delta\gamma}_1-\frac{\tilde\theta}{2(2f)}\tilde{\delta\pi}_1+\frac{2f\sqrt{2}}{12\tilde v}k^2\tilde{\delta\gamma}_2\right]  \\
	&&\ds- \tilde N\left(\frac{\sqrt{3}}{2}\tilde v^{1/3}\tilde V_{;I}  + \frac{\sqrt 3 f_{;I} \tilde v^{1/3}\tilde \theta^2}{16 f} - \frac{ f_{;I} }{2 \sqrt 3 f} \frac{k^2}{\tilde v^{1/3}} \right){\tilde Q}^I + \tilde N\frac{\sqrt{3}\tilde G^{IJ}\tilde \pi_J}{2\tilde v^{5/3}}\tilde P_I,
	\end{array}\right. \label{eq:CovDiff1} \\
	&&\left\{\begin{array}{rcl}
	\tilde D_\tau \tilde{\delta\gamma}_2&=&\ds-2\sqrt{\frac{2}{3}}\tilde v^{2/3}k\delta \tilde N_1+\frac{\tilde N}{2f}\left(4\tilde v^{1/3}\tilde{\delta\pi}_2+\tilde\theta\tilde{\delta\gamma}_2\right) \\
	\tilde D_\tau \tilde{\delta\pi}_2&=&\ds\sqrt{\frac{2}{3}}\tilde v^{1/3}k\delta \tilde N_1-\frac{2f}{\tilde v^{1/3}\sqrt{6}}k^2\delta \tilde N -\tilde N \frac{f_{;I}}{2\sqrt 6 f} {\tilde Q}^I\\
	&&\ds-\tilde N\left[\frac{1}{6\tilde v^{1/3}}\left(5\tilde\rho+3\tilde p-2f\frac{k^2}{2\tilde v^{2/3}}\right)\tilde{\delta\gamma}_2+\frac{\tilde\theta}{2f}\tilde{\delta\pi}_2+\frac{f\sqrt{2}}{6\tilde v}k^2\tilde{\delta\gamma}_1\right], 
	\end{array}\right. \label{eq:CovDiff2} \\
	&&\left\{\begin{array}{rcl}
	\tilde D_\tau {\tilde Q}^I&=&\ds\frac{1}{\tilde v}\tilde G^{IJ}\tilde \pi_J \delta \tilde N + \tilde N\left(\frac{1}{\tilde v}\tilde G^{IJ}\tilde P_J -\frac{\sqrt{3}}{2\tilde v^{5/3}}\tilde G^{IJ}\pi_J \tilde{\delta\gamma}_1+ \frac{\sqrt 3 \tilde G\uTsr \tilde f_{;J} \tilde v^{2/3}}{2 f} \tilde{\delta\pi}_1 \right)\\
	\tilde D_\tau \tilde P_I&=&\ds-\pi_I k\delta \tilde N_1-\tilde v \tilde V_{;I}\delta \tilde N -\tilde N \frac{f_{;I}}{2\sqrt 6 f} \frac{k^2}{\tilde v^{1/3}} \tilde{\delta\gamma}_2\\
	&&\ds-\tilde N\tilde v\left(\frac{k^2}{\tilde v^{2/3}}+\tilde V_{;IJ}-\frac{1}{\tilde v^2}\tilde R_I{}^{KL}{}_J\pi_K\pi_L  -\frac{3 \tilde v\tilde \theta^2}{8f}\left(\frac{f_{;I}f_{;J}}{f^2} - \frac{f_{;IJ}}{2 f}\right)\right){\tilde Q}^J\\
    &&-\tilde N \left(\frac{\sqrt{3}}{2}\tilde v^{1/3}\tilde V_{;I}  + \frac{\sqrt 3 f_{;I} \tilde v^{1/3}\tilde \theta^2}{16 f} - \frac{ f_{;I} }{2 \sqrt 3 f} \frac{k^2}{\tilde v^{1/3}}\right)\tilde{\delta\gamma}_1.
	 \end{array}\right. \label{eq:CovDiffQP}
\eea
These equations are to be solved on the surface of constraints, which read in this covariant formalism
\begin{equation}
    \tilde{\mathcal C^{(1)}} = 0\, , \quad \tilde{\mathcal D^{(1)}} = 0.
\end{equation}
Let us point out a few key differences between the equations of motion found here and their equivalent in the Einstein frame (Eqs. (3.1, 3.2$\&\,$3.3) in \cite{Grain_2026}). We will separate the terms in the right hand side of these equations in two categories. On the one hand, since we have chosen an arbitrary metric in the Jordan frame $\tilde G\uTsr$, we recover all the terms we can derive in the Einstein frame with the coupling metric $G\Tsr$, with the difference that any $\Mp^2/2$ has been replaced by a $f(\phi^I)$. On the other hand, we find additional terms in the coupling between the scalar field perturbations (or their covariant equivalent ${\tilde Q}^I$) with themselves and gravitational degrees of freedom. These \textit{new} couplings to $\tilde{\delta\gamma}_1$, $\tilde{\delta\pi}_1$ and the self coupling are non $k^2$ suppressed, whereas the one to $\tilde{\delta\gamma}_2$ and a second term coupling the field perturbation to $\tilde{\delta\gamma}_1$ are. We can also notice that the gradient terms appearing here that do not appear in the Einstein frame equations of motion all multiply configuration variables. 

\subsection{Separate universe}
We now develop cosmological perturbation theory around a homogeneous and isotropic background at scales larger than the Hubble radius. To this effect we work with the separate universe picture. We describe the universe at these scales as an ensemble of homogeneous and isotropic patches that, as a first approximation, evolve independently. These patches can be treated as independent Friedmann-Lemaître-Robertson-Walker (FLRW) universes. At the level of the perturbations themselves, this approach consists in introducing small deviations around the background solutions, which we assume to be homogeneous and isotropic, as proxies for the inhomogeneous and anisotropic cosmological perturbations at large scales. In practice, this means that we suppose $\delta\phi^I$ and $\tilde {\delta\pi}_J$ (or ${\tilde Q}^I$ and $\tilde P_J$) to no longer be space dependent. Similarly  $\tilde{\delta\gamma}_{ij}$ and $\tilde{\delta\pi}^{ij}$  can be expressed in terms of $\tilde{\delta v}$ and $\tilde{\delta\theta}$ which are also only time dependent. In this approach we also find that the anisotropic contributions to both $\tilde{\delta\gamma}\stsr$ and $\tilde{\delta\pi}\ustsr$ vanish \ie $\tilde {\delta\gamma}_2=\tilde {\delta\pi}_2=0$. We cannot a priori claim that the homogeneous and isotropic perturbations we define here are equal to those previously used. To avoid confusion we shall write these separate universe perturbations with a bar.

Once again we do not present the complete derivation of the separate universe hamiltonian and we refer the reader to App. A of \cite{Grain_2026} for more detail. There are two possible ways to derive this hamiltonian, on the one hand one can linearise the constraints as previously, with the major simplification that all gradient terms drop as a first approximation. On the other hand, since the perturbations obey the same symmetries as the background quantities, we can simply linearize the background equations of motion to get the equations of motions of the perturbations. We can then derive the hamiltonian from the equations of motion we obtain. We thus derive the following action
\bea
	\overline{S}^{(2)}_J=\ds\int\dd\tau\left[\overline{\tilde P}_I\dot{\overline{Q}}^I+\overline{\tilde{\delta\pi}}_1\dot{\overline{\tilde{\delta\gamma}}}_1-\overline{\delta N}\,\overline{\tilde{\mathcal{C}}}^{(1)}-2\tilde N\overline{\tilde{\mathcal{K}}}^{(2)}_\mathrm{cov}\right],
\eea
where the linear and second order constraints are
\begin{equation}
	\overline{\tilde{\mathcal{C}}}^{(1)}=\ds \left(\tilde v \tilde V_{;I} + \frac{3f_{;I} \tilde v \tilde \theta^2}{8f^2}\right)\overline {\tilde Q}^I+\frac{1}{\tilde v}\tilde G^{KI}\tilde \pi_K\overline {\tilde P}_I-\frac{\tilde v^{1/3}}{2\sqrt{3}}\left(\tilde\rho+3\tilde p\right)\overline{\delta\gamma}_1 -\frac{\sqrt{3}}{2 f}\tilde v^{2/3}\tilde \theta\,\overline{\delta\pi}_1 \,,\label{eq:first order scalar JSUA}
\end{equation}
\begin{align}
    \overline{\tilde{\mathcal{K}}}^{(2)}_{\mathrm{cov}} =& \frac{\tilde v}{4}\left(\tilde V_{;IJ} -\frac{1}{\tilde v^2}\tilde R_I{}^{KL}{}_J\bar\pi_K\bar\pi_L -\frac{3 \tilde v\tilde\theta^2}{8f}\left(\frac{f_{;I}f_{;J}}{f^2} - \frac{f_{;IJ}}{2 f}\right)\right)\left(\overline {\tilde Q}^I\overline {\tilde Q}^{J\,\star}+ \mathrm{c.c.}\right) \\
    &+\frac{1}{4\tilde v}\tilde G\uTsr\left(\overline {\tilde P}_I\overline {\tilde P}_J^\star+ \mathrm{c.c.}\right)  - \frac{\tilde v^{1/3}}{2f}|\overline{\tilde{\delta\pi}}_1|^2+\frac{1}{12\tilde v^{1/3}}\left(5\rho+3p\right)|\overline{\tilde{\delta\gamma}}_1|^2 \nonumber \\ 
	&-\frac{\tilde\theta}{8f}\left(\overline{\tilde{\delta\pi}}_1\overline{\tilde{\delta\gamma}}_1^\star + \mathrm{c.c.}\right)  +\left[ \frac{\sqrt{3}\tilde v^{1/3} \tilde V_{;I}}{4} + \frac{\sqrt 3 f_{;I} \tilde v^{1/3}\tilde \theta^2}{32 f} \right]\left( \overline {\tilde Q}^I\overline{\tilde{\delta\gamma_1}} ^\star+ \mathrm{c.c.} \right)\nonumber \\
    & + \frac{\sqrt 3 f_{;I} \tilde v^{2/3}}{4 f}\left( \overline {\tilde Q}^I \overline{\tilde{\delta\pi}}_1 ^\star+ \mathrm{c.c.} \right) -\frac{\sqrt{3} \tilde G\uTsr\tilde \pi_J}{4 \tilde v^{5/3}}\left(\overline {\tilde P}_I\overline{\tilde{\delta\gamma}}_1 ^\star+ \mathrm{c.c.}\right)\nonumber.
\end{align}
We have already written everything in a manifestly covariant manner here. The resulting equations of motion for the isotropic degrees of freedom are
\bea
	&&\left\{\begin{array}{rcl}
	D_\tau \overline{\tilde{\delta\gamma}}_1&=&
	-\frac{\sqrt{3}}{2f}\tilde v^{2/3}\tilde\theta\,\overline{\delta \tilde N} -\frac{\tilde N}{2f}\left(2\tilde v^{1/3}\overline{\tilde{\delta\pi}}_1+\frac{\tilde\theta}{2}\overline{\tilde{\delta\gamma}}_1\right) + \frac{\sqrt 3 \tilde f_{;I} \tilde v^{2/3}}{2 f}\,\overline {\tilde Q}^I\\[1ex]
	D_\tau \overline{\tilde{\delta\pi}}_1&=& \frac{\tilde v^{1/3}}{\sqrt{3}}\frac{1}{2}\left(\tilde\rho+3\tilde p\right)\overline{\delta \tilde N} -\tilde N\left[\frac{1}{6\tilde v^{1/3}}\left(5\tilde\rho+3\tilde p\right)\overline{\tilde{\delta\gamma}}_1-\frac{\tilde\theta}{4f}\overline{\tilde{\delta\pi}}_1\right]  \\
	&&\ds- \tilde N\left(\frac{\sqrt{3}}{2}\tilde v^{1/3}\tilde V_{;I} + \frac{\sqrt 3 \tilde f_{;I} \tilde v^{1/3}\tilde \theta^2}{16 f}\right)\overline {\tilde Q}^I + \tilde N\frac{\sqrt{3}\tilde G^{IJ}\tilde \pi_J}{2\tilde v^{5/3}}\overline{\tilde  P}_I, 
    \end{array}\right. \label{eq:CovDiff1SU} \\
	&&\left\{\begin{array}{rcl}
	D_\tau \overline {\tilde Q}^I&=&\ds\frac{1}{\tilde v}\tilde G^{IJ}\pi_J \,\overline{\delta \tilde N}+ \tilde N\left(\frac{1}{\tilde v}\tilde G^{IJ}\overline P_J-\frac{\sqrt{3}}{2\tilde v^{5/3}}\tilde G^{IJ}\pi_J \overline{\tilde{\delta\gamma}}_1+ \frac{\sqrt 3 \tilde G\uTsr \tilde f_{;J} \tilde v^{2/3}}{2 f}\overline{\tilde{\delta\pi}}_1\right)\\
	D_\tau \overline{\tilde  P}_I&=&\ds-\tilde v \tilde V_{;I}\,\overline{\delta \tilde N} -\tilde N \left(\frac{\sqrt{3}}{2}\tilde v^{1/3}\tilde V_{;I}+ \frac{\sqrt 3 \tilde f_{;I} \tilde v^{1/3}\tilde \theta^2}{16 f}\right)\overline{\tilde{\delta\gamma}}_1 \\
	&&\ds-\tilde N\tilde v\left(\tilde V_{;IJ}-\frac{1}{\tilde v^2}\tilde R_I{}^{KL}{}_J\pi_K\pi_L-\frac{3 \tilde v\tilde \theta^2}{8f}\left(\frac{\tilde f_{;I}\tilde f_{;J}}{f^2}- \frac{\tilde f_{;IJ}}{2 f}\right)\right)\overline {\tilde Q}^J.
	 \end{array}\right. \label{eq:CovDiffQPSU}
\eea
which once again need to be solved on the surface of constraints. In this SUA picture, the surface of constraint is one dimensional and defined by $\overline{\tilde{\mathcal{C}}}^{(0)} = 0$.

\subsection{Frame dependent cut-off scale}
\subsubsection*{Jordan frame conditions}\label{ssec:Jordan frame conditions}
We are now in a position to derive the validity conditions of this approximation. We wish to verify whether the equations of motions in Eq. (\ref{eq:CovDiffQPSU}) accurately give the dynamics of the equations of motion in Eq. (\ref{eq:CovDiffQP}). This will be the case when the gradients are neglected in the full perturbation theory. We derive here the conditions at the level of the constraints and equations of motion and compare them to those derived in the Einstein frame.

Let us first briefly discuss the case of the diffeomorphism constraint. Since it is lost in the separate universe approach and a pure gradient in the full perturbation approach, one could argue that neglecting all gradient terms automatically makes these two match. However the reality is slightly more subtle than this. Since the diffeomorphism constraint is not affected by the conformal transform between the Einstein frame and the Jordan frame, our argument is identical to that presented in \cite{Grain_2026}. We can introduce a fictitious diffeomorphism constraint in the separate universe approximation $\overline{\tilde{\mathcal D}}^{(1)}$, which we show to be a constant provided $\overline{\tilde{\mathcal C}}^{(1)} = 0$. In practice this fictitious constraint is built by taking the diffeomorphism constraint in the full picture Eq. (\ref{eq:constcov}), discarding the anisotropic contributions, and replacing the isotropic ones with their separate universe proxies. The mismatch between this fictitious constraint and the \textit{real} one in the complete perturbation theory approach is small and $k^2$-suppressed, thus insuring that it remains small after a time evolution through the separate universe dynamics.

The condition we derive thus comes from comparing the first order scalar constraint in the cosmological perturbation theory approach Eq. (\ref{eq:constcov}) and in the separate universe picture Eq. (\ref{eq:first order scalar JSUA}). We find that there are two gradient terms in the latter, however one multiplies $\tilde{\delta\gamma}_2$ which is set to vanish in order to compare both methods. Neglecting the remaining gradient terms thus simply reads
\begin{equation}
    \frac{k^2}{\tilde v^{2/3}} \ll \frac{1}{2|f|}\left|\frac{\tilde G\uTsr\tilde \pi_I\tilde \pi_J}{\tilde v^2} + \tilde V\right|. \label{eq:JF condition 1}
\end{equation}
This condition is very close to the one derived in the Einstein frame, and identical if we simply take $f(\phi^I) = \frac{\Mp^2}{2}$. Let us now aim to neglect gradients in $\tilde{\mathcal K}_{\mathrm{\text{cov}}}^{(2)}$, which will lead to several sets of conditions. As mentioned previously, gradients only arise through quadratic terms involving the configuration variables: $\left({\tilde Q}^I,\tilde{\delta\gamma}_1,\tilde{\delta\gamma}_2\right)\tilde{\mathcal M}\left({\tilde Q}^J,\tilde{\delta\gamma}_1,\tilde{\delta\gamma}_2\right)^\dagger$.We call $\tilde{\mathcal M}$ the mass matrix.
\beq
	\tilde{\mathcal{M}} \equiv \left(\begin{array}{ccc}
		\frac{\tilde v}{4}\tilde M\Tsr & \tilde L_I - \frac{ f_{;I} }{8 \sqrt 3 f} \frac{k^2}{\tilde v^{1/3}} & \frac{f_{;I}}{8\sqrt 6 f} \frac{k^2}{\tilde v^{1/3}} \\
		 \tilde L_I^T - \frac{ f_{;I}^\mathrm{T} }{8 \sqrt 3 f} \frac{k^2}{\tilde v^{1/3}} & \frac{1}{\tilde v^{1/3}}\left(\frac{\tilde G\uTsr\tilde \pi_I\tilde \pi_J}{\tilde v^2}+\frac{\tilde V}{2}-\frac{f\,k^2}{2\tilde v^{2/3}}\right) & \frac{\sqrt{2}fk^2}{12\tilde v} \\
		 \frac{f_{;I}^{\mathrm T}}{8\sqrt 6 f} \frac{k^2}{\tilde v^{1/3}} & \frac{\sqrt{2}f\,k^2}{12\tilde v} & \frac{1}{\tilde v^{1/3}}\left(\frac{\tilde G\uTsr\tilde \pi_I\tilde \pi_J}{\tilde v^2}+\frac{\tilde V}{2}-\frac{f\,k^2}{4\tilde v^{2/3}}\right)
	\end{array}\right),
\eeq
with 
\begin{align}
    \tilde L_I \equiv& \frac{\sqrt{3}\tilde v^{1/3}}{8}\tilde V_{;I}  + \frac{\sqrt 3 f_{;I} \tilde v^{1/3}\tilde \theta^2}{64 f}\,, \\
    \tilde M\Tsr \equiv&\frac{k^2}{\tilde v^{2/3}}\tilde G_{IJ} + \tilde V_{;IJ} -\tilde{\mathcal{R}}_{IJ} -\frac{3 \tilde \theta^2}{4f}\left(\frac{\tilde f_{;I}\tilde f_{;J}}{f^2}- \frac{\tilde f_{;IJ}}{2 f}\right)\,,\\
    \tilde{\mathcal{R}}_{IJ}\equiv& \frac 1 {\tilde v ^2}\tilde R_I{}^{KL}{}_J\tilde \pi_K\tilde \pi_L.
\end{align}
In all generality, in order to properly neglect gradient terms in this matrix we should diagonalise it and neglect the gradients in the eigenvalues. This will give us a set of at most $n+2$ conditions, with $n$ the number of fields. However we can also neglect the gradients directly, which will give stronger conditions than necessary but correct ones nonetheless. In order to do so, we can start by considering the top left corner which involves the self couplings between the field perturbations $\tilde M\Tsr$. Since this matrix is symmetric and real it can be diagonalised. We call $\tilde m_I^2$ its eigenvalues, with $I$ running from $1$ to $n$. Neglecting the gradients in said diagonal block boils down to
\beq
    \frac{k^2}{\tilde v^{2/3}} \ll \tilde m{_I}^2. \label{eq:JF condition 3}
\eeq
Turning to the bottom right corner of our mass matrix we find two identical conditions that read 
\begin{equation}
    \frac{k^2}{\tilde v^{2/3}} \ll \frac{1}{2|f|}\left|\frac{\tilde G\uTsr\tilde \pi_I\tilde \pi_J}{\tilde v^2} + \frac{\tilde V}{2}\right|. \label{eq:JF condition 2}
\end{equation}
If we assume slow roll, this condition becomes equivalent to the one derived from the first order constraint\footnote{Up to a dimensionless prefactor of order one.}. Finally, we can very easily impose to neglect the gradient term in the ${\tilde Q}^I \tilde{\delta\gamma}_1$ coupling, which leads to
\begin{equation}
    \left|\frac{ f_{;I} }{f}\right| \frac{k^2}{\tilde v^{2/3}} \ll \left|3\tilde V_{;I}  + \frac{ 3 f_{;I} \tilde \theta^2}{8 f}\right|. \label{eq:JF condition 4}
\end{equation}
We are now left with the off diagonal terms containing double gradients only. Let us call $\tilde{\mathcal P}$ the matrix containing these terms, and $\tilde{\mathcal O} = \tilde {\mathcal M} - \tilde{\mathcal P}$. The trace norm of $\tilde{\mathcal P}$ is proportional to $k^4/v^{4/3}$, whereas the trace norm of $\tilde{\mathcal O}$ is proportional to products of the right hand sides of Eqs. (\ref{eq:JF condition 1}, \ref{eq:JF condition 2} $\&$ \ref{eq:JF condition 3}), which we have all shown to be much greater to $k^2/v^{2/3}$. These additional double gradient terms can thus be neglected. These conditions are all sufficient but not necessary, it is clear that a complete diagonalization of the matrix would allow us to neglect the gradients in both of these couplings at once, just like for the purely gravitational block.

\subsubsection*{Einstein frame conditions}\label{sssec:Einstein frame conditions}
We have finally arrived at a simple set of conditions the coarse graining scale needs to verify in order to capture the large scale dynamics accurately with the separate universe approach. Let us now compare this set of conditions with those derived in the Einstein frame. We will not re-derive them as they have been presented at length in \cite{Grain_2026}. We provide here the main results. The first order constraint from which one set of condition is derived reads
\begin{equation}
    \mathcal C^{(1)} =  vV_{;I}\, Q^I+\frac{1}{v}G^{KI}\pi_KP_I-\frac{v^{1/3}}{\sqrt{3}}\left[\frac{1}{2}\left(\rho+3p\right)+\Mp^2\frac{k^2}{v^{2/3}}\right]\delta\gamma_1+\frac{\Mp^2}{\sqrt{6}}\frac{k^2}{v^{1/3}}\delta\gamma_2 -\frac{\sqrt{3}}{\Mp^2}v^{2/3}\theta\,\delta\pi_1 .
\end{equation}
And the mass matrix that appears in this frame is
\beq
	\mathcal{M}=\left(\begin{array}{ccc}
		\frac{v}{4}\left(\frac{k^2}{v^{2/3}} G_{IJ} + V_{;IJ} -\mathcal{R}_{IJ}\right) & \frac{\sqrt{3}v^{1/3}}{8}V_{;I} & 0 \\
		 \frac{\sqrt{3}v^{1/3}}{8}V^\mathrm{T}_{;I} & \frac{1}{v^{1/3}}\left(\frac{\pi^2_\sigma}{v^2}+\frac{V}{2}-\frac{\Mp^2k^2}{4v^{2/3}}\right) & \frac{\sqrt{2}\Mp^2k^2}{24v} \\
		 0 & \frac{\sqrt{2}\Mp^2k^2}{24v} & \frac{1}{v^{1/3}}\left(\frac{\pi^2_\sigma}{v^2}+\frac{V}{2}-\frac{\Mp^2k^2}{8v^{2/3}}\right)
	\end{array}\right), \label{eq: mass matrix einstein}
\eeq
where all the quantities have similar definitions to those with a tilde in the Jordan frame. We can then diagonalise this matrix by block to get two sets of conditions as done in this work for the Jordan frame. Let us call $m_I$ the $n$ eigenvalues of the top left block. The final set of conditions is thus:
\begin{align}
    \frac{k^2}{ v^{2/3}} \ll& \frac{1}{2\Mp^2}\left| \frac{ G\uTsr\pi_I\pi_J}{ v^2} +  V\right|\,, \label{eq:EF condition 1}\\
    \frac{k^2}{ v^{2/3}} \ll& \frac{1}{\Mp^2}\left|\frac{ G\uTsr\pi_I\pi_J}{ v^2} + \frac{ V}{2}\right|\, , \label{eq:EF condition 2}\\
    \frac{k^2}{ v^{2/3}} \ll&  m{_I}^2. \label{eq:EF condition 3}
\end{align}
We can immediately note that their is one less set of conditions in this frame.  We can however notice that in this frame too, the first two conditions (obtained from the first order constraint and from the gravitational block) match if we assume slow roll. Let us show that these conditions can't be matched one-to-one. We focus on the conditions coming from the gravitational block in the Einstein frame Eq. (\ref{eq:EF condition 2}), and input the replacements defined by the conformal transform Eqs. (\ref{eq:G E & J}$\&\,$\ref{eq:pi E & J}). Considering the first term only, this gives
\begin{equation}
    G\uTsr\pi_I\pi_J \to \left[\frac{\Mp^2}{2f} \tilde G_{IJ} + \frac{3\Mp^2}{2f^2}f_{,I}f_{,J}\right] \left(\tilde\pi_I + \frac 3 {f \tilde v} \tilde G^{KL}f_{,I}f_{,L}\tilde\pi_K\right)\left(\tilde\pi_J + \frac 3 {f \tilde v} \tilde G^{MN}f_{,M}f_{,J}\tilde\pi_N\right). \label{eq:condition comparison}
\end{equation}
We immediately see that we do not recover the conditions appearing in the Jordan frame Eqs. (\ref{eq:JF condition 1}, \ref{eq:JF condition 2}, \ref{eq:JF condition 3}$\&\,$\ref{eq:JF condition 4}). On the one hand no double derivative of $f$ appear in this redefinition, making the matching to the the Jordan frame field mass conditions impossible. On the other hand, the square of simple derivatives that do appear make the matching to the condition from the gravitational block and off diagonal terms just as problematic.

\section{Applications to a Higgs like toy model}\label{sec:Higgs}
Let us introduce a Higgs like toy model in which to compute the mass matrices in both frames. Let us call $\phi = h e^{i\alpha}$ a complex scalar field, with the following action:
\bea
    S = \int \dd^4x \sqrt{-\tilde g} \left[(\frac{\Mp^2}{2} + \frac{\xi h^2}{2})  \tilde{R} - \frac 1 2 \partial_\mu h\partial^\mu h -  - \frac {h^2} 2 \partial_\mu \alpha\partial^\mu \alpha - \frac \lambda 4 (h^2-h_v^2)^2\right]
\eea
where $\xi$ and $\lambda$ are parameters of our model, and $h_v$ is the vacuum expectation value of the radial field. We can notice here that the coupling metric, the potential and the non minimal coupling to gravity all depend only on the modulus of our field $h$ and not its phase $\alpha$. We can notice that this action is written in the Jordan frame, and we can proceed to the conformal transform to go to the Einstein frame. In this case, we have $\Omega^2(h) = 1 + \frac{\xi h^2}{\Mp^2}$, or equivalently $f \equiv \frac{\Mp^2}{2} + \frac{\xi h^2}{2}$ \cite{BEZRUKOV2008703}. We also proceed to a field redefinition of the modulus of our field, and study $\chi$ defined as
\bea
    \frac{\dd \chi}{\dd h} = \sqrt{\frac{\Omega^2 + 6 \xi^2h^2/\Mp^2}{\Omega^4}}.
\eea
Once this is done, two major simplifications are possible. We can either study the small field limit (\ie $h \ll \Mp/\sqrt \xi$ or equivalently $\chi \ll \sqrt 6 \Mp$ \cite{Bezrukov_2014, R_s_nen_2019, Garc_a_Bellido_2009}) or the large field limit (\ie $h \gg \Mp/\sqrt \xi$ or equivalently $\chi \gg \sqrt 6 \Mp$ \cite{BEZRUKOV2008703, Barb_n_2009, Lerner_2010, Kaiser_2014, Kaiser_2013}). We focus in this work on the latter, in which our new coupling metric is defined as
\beq
    G_{IJ} = \left(\begin{array}{cc}
        1 & 0 \\
        0 & \frac{\Mp^2}{\xi}e^{-\sqrt\frac 2 3 \frac \chi\Mp}
    \end{array}\right),
\eeq
and the potential reads
\beq
    V(\chi,\alpha) = \frac{\lambda\Mp^2}{2\xi}\left(1-e^{-\sqrt\frac 2 3 \frac \chi\Mp}\right)^{2}. \label{eq:potential higgs EF}
\eeq
This expression is precisely what we have in Starobinsky inflation \cite{STAROBINSKY198099}. We can notice that neither the coupling metric nor the potential depend on the field $\alpha$. From this action we can write the background equations of motion of both field $\left(\chi,\alpha\right)$, and their respective momenta $\left(\pi_\chi,\pi_\alpha\right)$
\begin{eqnarray}
    &&\left\{\begin{array}{rcl}
       \chi' &=& H\frac{\pi_\chi}{v} \\
        \pi_\chi' &=& -v H V_{,\chi} - \frac{G^{\alpha\alpha}_{,\chi}}{2v} H\pi_\alpha^2 \label{eq:Eom pi chi}
    \end{array}\right.\\
    &&\left\{\begin{array}{rcl}
       \alpha' &=& \frac{\Mp^2\pi_\alpha}{\xi v}He^{\sqrt{\frac 2 3}\frac{\chi}{\Mp}} \\
        \pi_\alpha' &=& 0 \label{eq:Eom pi alpha}
    \end{array}\right.
\end{eqnarray}
where we take the number of e-folds $\N = \text{ln}(a)$ as a time coordinate. Let us introduce another useful quantity for clarity
\begin{equation}
    Y = e^{-\sqrt{\frac 2 3}\frac{\chi}{\Mp}}\,,
\end{equation}
which is a small parameter in the large field limit.

\subsection{Background dynamics}
We set on solving the background equations of motion. We start by assuming a perfectly radial trajectory before checking that this assumption is reasonable and does not break due to quantum fluctuations along the polar direction.
\subsubsection*{Purely radial trajectory}
The momenta $\pi_\alpha$ is a constant of motion in this picture, and this is the case in both frames since Eq. (\ref{eq:pi E & J}) shows that $\tilde\pi_\alpha = \pi_\alpha$. Let us initially suppose that it is vanishing. Computing the equation of motion for the field $\chi$ now leads to the usual Klein-Gordon equation
\begin{equation}
    \chi'' + (3-\epsilon) \chi' + \frac{V_{;\chi}}{H^2} = 0\,,
\end{equation}
where $\epsilon$ is the first slow roll parameter. Assuming slow roll ($\epsilon \ll 1$) allows us to use the relation $H^2 \simeq \frac{V}{3\Mp^2}$. We also use the second slow roll condition to discard $\chi''$. Reinjecting these leads to 
\begin{equation}
    \chi' \simeq -2\sqrt{\frac 2 3}\Mp Y.
\end{equation}
Where we have used the fact that $Y \ll 1$. We can finally solve this to
\begin{equation}
    \chi_0(\N) = \sqrt{\frac 3 2} \Mp \text{ln}\left(-2\sqrt{\frac 2 3 } (\N_i-\N)\right)\,,
\end{equation}
with $\N_i$ the initial time\footnote{And in general and index $_i$ indicates initial conditions.}. From this we can compute the rest of our variables, for example at leading order in $Y$
\begin{equation}
    \pi_\chi = -\frac{e^{3\N}\sqrt\lambda \Mp}{\sqrt 3 \xi}Y.
\end{equation}
\subsubsection*{Introducing small polar perturbation}
Let us now reintroduce a small polar momentul $\pi_\alpha$ perturbatively. To this effect we assume that $\pi_\alpha$ is non vanishing but small enough to verify 
\begin{equation}
    \left|\frac{G^{\alpha\alpha}_{,\chi}}{2v}\pi_\alpha^2\right|\ll \left|v V_{,\chi} \right|. \label{eq:condition pi alpha}
\end{equation}
This condition comes from neglecting the $\pi_\alpha$ contributions in Eq. (\ref{eq:Eom pi chi}), so it is the second slow roll condition. We shall also verify that this conditions holds through the evolution. Let us write $\chi = \chi_0 + \delta\chi$, where $\chi_0$ is the solution we just derived assuming $\pi_\alpha = 0$. Keeping the slow roll assumption, the equation of motion for $\delta\chi$ reads
\begin{equation}
    3 \delta \chi'= -\frac{G^{\alpha\alpha}_{,\chi}}{2v^2 H} \pi_\alpha^2.
\end{equation}
and solves to
\begin{align}
    \delta\chi(\N) = \frac{2 \xi^3}{27\lambda \Mp^5}\pi_\alpha^2e^{-6\N}\left(6(\N_i-\N) - 1\right).
\end{align}
We can already note the this correction is not only suppressed by $\pi_\alpha^2$, it is also suppressed by $e^{-6\N}$. It is thus safe to assume that these perturbations are indeed small. However let us verify that our assumption on $\pi_\alpha$ remains true throughout evolution. Reinjecting our solution $\chi_0$ in Eq. (\ref{eq:condition pi alpha}), we get 
\begin{equation}
    \pi_\alpha^2 \ll \frac{\lambda \Mp^5}{2\sqrt 3 \xi^3}e^{6\N}Y_0^2.
\end{equation}
Considering the shape of the potential, during evolution (so as $\N$ increases), $\chi$ decreases. The strictest condition for $\pi_\alpha$ to satisfy is thus
\begin{equation}
    \pi_\alpha^2 \ll \frac{\lambda \Mp^5}{2\sqrt 3 \xi^3}e^{6\N_i}e^{-2\sqrt{\frac 2 3}\frac{\chi_i}{\Mp}}. \label{eq: condition pi alpha}
\end{equation}
\subsubsection*{Polar momentum quantum fluctuation}
Let us compute the power spectrum of $\delta\pi_\alpha$ to check whether the condition we just found can hold or if quantum fluctuations along the polar momentum can break it. In the Lagrangian we are working with, the kinetic term for the $\alpha$ direction reads $\mathcal L \supset \frac 1 2 G_{\alpha\alpha}(\partial\alpha)^2$. We can define the canonical variable 
\begin{equation}
    s \equiv \sqrt{G_{\alpha\alpha}}\alpha \simeq e^{-\frac{\chi}{\sqrt 6 \Mp}}.
\end{equation}
We call the perturbation of this variable $Q_s$\footnote{This notation is consistent with the way the Mukhanov Sasaki variables are usually denoted since we are actually studying the entropic Mukhanov Sasaki variable in the flat gauge.}. This is a light scalar field in an approximate de Sitter space, its vacuum expectation value thus reads
\begin{equation}
    \braket{Q_s^2} = \left(\frac{H}{2\pi}\right)^2.
\end{equation}
From this we can go back to the original field perturbation 
\begin{equation}
    \delta\alpha = \frac{H}{2\pi\sqrt{G_{\alpha\alpha}}}.
\end{equation}
To get the perturbation of the associated momenta, we remind the reader that in the separate universe approach $\pi_\alpha = vG_{\alpha\alpha}\dot\alpha$. Expanding this around a purely radial trajectory, we get
\begin{equation}
    \delta\pi_\alpha \simeq \frac{H^2}{2\pi}e^{3\N}\frac{\Mp}{\sqrt\xi}e^{-\frac{\chi}{\sqrt 6 \Mp}}.
\end{equation}
Finally the power spectrum is 
\begin{equation}
    \mathcal P_{\pi_\alpha} = \frac{H^4}{4\pi^2}\frac{\Mp^2}{\xi}e^{6N-\sqrt{\frac 2 3}\frac{\chi}{\Mp}}.
\end{equation}
This allows us to compare the typical fluctuation amplitudes to the bound we previously derived Eq. (\ref{eq: condition pi alpha}). 
\begin{equation}
    \delta\pi_\alpha \ll \pi_\alpha^\text{max} \, \Leftrightarrow \, e^{\frac 1 {\sqrt 6}\frac{\chi_i}{\Mp}} \ll \frac{24\pi\xi}{\lambda}.
\end{equation}
The explicit time dependence drops in this comparison, and we are left with a condition on the initial field value. To evaluate this let us recall some typical results of both Higgs inflation and Starobinsky inflation models. We take the usual estimate of $ \xi  \simeq 10^4$ and $\lambda \simeq 10^{-1}$ \cite{Lerner_2010, drees2021overshootingcriticalhiggsinflation}, and write that the total number of e-folds is roughly given by $\N \simeq \frac 3 4 e^{\sqrt{\frac 2 3}\frac{\chi_i}{\Mp}}$. Taking $\N = 60$ thus leads to $80^{1/2}\ll10^5$ which is perfectly reasonable. We do not expect the quantum fluctuations along the polar momentum to lead us out of the validity domain for our assumption on $\pi_\alpha$, hence we can assume the we stay on the slow roll radial trajectory.

\subsection{Mass matrix and validity conditions in the Einstein frame}
Now that we have shown that the vanishing $\pi_\alpha$ case does not easily get violated, let us compute the mass matrix Eq. (\ref{eq: mass matrix einstein}) in this case and derive the validity conditions for the separate universe approach. As mentioned previously the ideal case would be if the complete mass matrix was written in its diagonal form, which we are guaranteed exists since it is a real and symmetric matrix. However let us keep the more naive approach and look at the conditions in each diagonal block.

The upper left block is already diagonal in this case, at leading order in $Y$ it reads
\begin{equation}
    \frac{k^2}{v^{2/3}} G_{IJ} + V_{;IJ} -\mathcal{R}_{IJ} = \left(\begin{array}{cc}
		\frac{k^2}{v^{2/3}} -\frac{\lambda \Mp^2}{3\xi^2}Y & 0  \\
		 0 & \frac{k^2}{v^{2/3}}\frac{\Mp^2}{\xi}Y + \frac{\lambda\Mp^4}{6\xi^3}Y^2
	\end{array}\right).
\end{equation}
Let us note that in this particular case, terms coming from the Ricci tensor are subdominant and do not contribute to the conditions we derive. Neglecting gradients in this diagonal block, as in Eq. (\ref{eq:EF condition 3}), leads to a single\footnote{Up to a constant prefactor of order 1.} condition 
\begin{equation}
    \frac{k^2}{v^{2/3}} \ll \frac{\lambda \Mp^2 Y}{4\xi^2}.
\end{equation}
Let us focus now on the conditions involving only gravitational perturbations Eq. (\ref{eq:EF condition 2}). Injecting our results for both the momenta and the potential to the general expression of the condition, we get 
\begin{equation}
    \frac{k^2}{v^{2/3}}\ll \frac{\lambda \Mp^2}{4\xi^2}.
\end{equation}
Let us note that this is also the condition we get from the first order constraint Eq. (\ref{eq:EF condition 1}), as we have previously shown that these two conditions are equivalent in slow roll inflation regardless of the model we are considering. Let us discuss the conditions we obtained. We naturally have less than four conditions coming from the mass matrix since the opposite would imply that we need to properly compute the eigenvalues and neglect the gradients there. And we actually only have two conditions in total. However by definition, we have $Y\ll1$, which makes the first inequality we derived much more stringent, and thus that is the one which needs to be verified in order to justify the use of the separate universe approach. We thus found that the lightest mass leads to the most stringent condition, which is consistent with formal results derived in \cite{Artigas_2022} and \cite{Grain_2026}.

\subsection{Mass matrix and validity conditions in the Jordan Frame}

Let us proceed similarly in the Jordan frame and derive the validity conditions for the separate universe approach. Since all the previous computations have been derived in the Einstein frame, to get the expressions in the Jordan frame we can simply go back to the definition of the canonical field $\chi$ and proceed to the inverse transformation. The most important quantity we need is the conjugate momentum to the field $h$. It reads in the large field limit and at leading order in $h$
\begin{equation}
    \pi_h = -\frac{\tilde v\sqrt \lambda \Mp}{3\sqrt 2 \xi h}.
\end{equation}
We have checked that once we translated our resolution back to the Jordan frame, the potential remains the dominant term in both the pressure and the momentum. Thus we can assume slow roll in this frame too. Let us focus first on the condition coming from the first order constraint Eq. (\ref{eq:JF condition 1}). Plugging in the expression of the potential, we get 
\begin{equation}
    \frac{k^2}{\tilde v^{2/3}} \ll \frac{\lambda}{8\xi}h^2.
\end{equation}
Let us now focus on the conditions coming from the mass matrix. Since $h$ is a more natural field to use in the Jordan frame we keep it as such and the conditions we will derive will depend on it and not the canonical Einstein frame field $\chi$. As previously in this frame we treat the conditions element by element. The top left corner reads
\begin{equation}
    \frac{k^2}{\tilde v^{2/3}}\tilde G_{IJ} + \tilde V_{;IJ} -\tilde{\mathcal{R}}_{IJ} -\frac{3 \tilde \theta^2}{2f}\left(\frac{\tilde f_{;I}\tilde f_{;J}}{f^2}- \frac{\tilde f_{;IJ}}{2 f}\right) = \left(\begin{array}{cc}
		\frac{k^2}{v^{2/3}} +\frac{3 \lambda }{2}h^2 & 0  \\
		 0 & \frac{k^2}{v^{2/3}}h^2 - \frac{\lambda}{2}h^4
	\end{array}\right).
\end{equation}
In this case the Ricci scalar vanishes completely, and the two other terms contribute at the leading order. This block is already diagonal so we only need to read off the conditions and we get a unique (up to an order $1$ prefactor) condition, instead of the two distinct conditions expected in Eq. (\ref{eq:JF condition 3}), it reads
\begin{equation}
    \frac{k^2}{\tilde v^{2/3}} \ll \lambda h^2.
\end{equation}
Turning towards the gravitational block Eq. (\ref{eq:JF condition 2}), we can derive the condition exactly as we did for the Einstein frame and we get 
\begin{equation}
    \frac{k^2}{\tilde v^{2/3}} \ll \frac{\lambda h^2}{\xi}.
\end{equation}
Let us finally turn to the set of conditions coming from the off diagonal block Eq. (\ref{eq:JF condition 4}). We get at linear order in $h$
\begin{equation}
    \frac{\sqrt{3}\tilde v^{1/3}}{8}\tilde V^\mathrm{T}_{;I}  + \frac{\sqrt 3 f_{;I}^\mathrm{T} \tilde v^{1/3}\tilde \theta^2}{64 f} - \frac{ f_{;I}^\mathrm{T} }{8 \sqrt 3 f} \frac{k^2}{\tilde v^{1/3}} = \begin{pmatrix} \frac{e^\N \lambda \xi h^5}{32\sqrt 3} - \frac{k^2e^{-\N}}{4\sqrt 3 h} \\ 0 \end{pmatrix}.
\end{equation}
This leads to the final condition
\begin{equation}
    \frac{k^2}{\tilde v^{2/3}}\ll\frac{\lambda\xi h^6}{8} .
\end{equation}
This condition is much less stringent than the previous two since the right hand term is not only enhanced by multiple powers of $h$, it is also enhanced by $\xi$, both being large parameters in our problem. Thus the hardest condition the verify is the one obtained in both the first order constraint and the gravitational block. Let us point out that in this particular example, the \textit{new} condition coming from the Jordan Frame is not relevant since it is much less stringent. This cannot however be generalised to any case. 

We can once again check that there is no one-to-one correspondence between the conditions derived in one frame or the other. Let us note a key difference. We found on the one hand that in the Einstein frame the most stringent constraint came from the effective mass of the radial direction. On the other hand, in the Jordan frame we found that it is the condition coming from the gravitational sector that is the hardest to verify. Finally, taking typical values for Higgs inflation ($\lambda \simeq 0.1$ and $\xi \simeq 10^4$), and evaluating the large field limit to $h\ge 10^{16}\mathrm{GeV}$ (or $Y\simeq10^{-2}$), the most stringent condition can be evaluated to 
\begin{equation}
    \frac{k^2}{\tilde v^{2/3}} \ll 10^{25}\,,\quad \frac{k^2}{v^{2/3}} \ll 10^{23}.
\end{equation}
The first condition is for the Jordan frame, and the second for the Einstein frame. We find similar results despite the conditions coming from two difference types of couplings, however we have seen in the discussion following Eq. (\ref{eq:condition comparison}) that the conditions cannot be matched one-to-one. It is thus quite remarkable to have found a similar numerical evaluation for both.

\section{Conclusion}\label{sec:conclusion}

In this paper we have explored the Hamiltonian description of cosmological perturbation theory and of the separate universe approach for multifield models of inflation. We have extended previous work by allowing the multiple scalar fields to be non minimally coupled to gravity. In this context we presented the usual conformal transformation allowing us to go from the Jordan frame to the Einstein frame, in which the non minimal coupling to gravity is absorbed by the coupling between the fields. The separate universe approach that we studied consists in discarding all anisotropies, including those in the scalar perturbations. We compared the dynamics of the scalar perturbation in a general construction taken to large scale to the dynamics derived with the separate universe approach in order to get them to match.

Our conclusion can be found in Sec. \ref{ssec:Jordan frame conditions}, and states that in an arbitrary gauge, the separate universe approach matches cosmological perturbation theory as long as two distinct conditions are met. We need to respect a set of equations on the scales, which state that we need to be working at large enough scales, compared to the horizon scale and the effective masses of all the fields, as well as a third condition coming from the non minimal coupling to gravity. The second condition imposes the anisotropic degrees of freedom to be set to vanish, since they are discarded in the separate universe approach.

Since the number of conditions we found on the scales was greater in this case than in the classically equivalent Einstein frame, we applied these methods to a Higgs-like toy model of inflation in Sec. \ref{sec:Higgs} in order to compare the set of conditions directly. We studied this toy model in the so-called large field limit in which many simplifications occur, as well as assuming slow roll. We also originally assumed a perfectly radial trajectory before introducing a small polar deviation, which we proved to be a reasonable assumption.  We computed the mass matrices to derive the conditions on the scales in both frames. We found that the conditions derived in both frames cannot be matched one-to-one, and that the most stringent condition does not come from the same type of couplings in both frames. In the Einstein frame case it is the lightest effective field mass that gives the tightest condition, whereas in the Jordan frame it is the self coupling of gravity that give the tightest constraints.

This paper partially answers a question that was raised in previous works about how the separate universe approach is affected by the conformal transform and non minimal couplings to gravity. Some aspects still remain to be explored, we have stuck to a gauge independent description here but we have seen that in order to truly match the separate universe approach to large scale perturbation theory the perturbed Lagrange multipliers need to match, which can only be checked after fixing the gauge \cite{Grain_2026}. We could check whether different frames lead to different conditions coming from the gauge fixing procedure, we however leave this for future work.
\bibliographystyle{JHEP}
\bibliography{ref}

\end{document}